\documentclass[11pt]{article} 

\usepackage[T1]{fontenc}
\usepackage[utf8]{inputenc}
\usepackage{authblk}

\usepackage[english]{babel} 
\usepackage[T1]{fontenc}
\usepackage{amsmath,mleftright}
\usepackage{amsfonts} 
\usepackage{xparse}

\NewDocumentCommand{\evalat}{sO{\big}mm}{%
  \IfBooleanTF{#1}
   {\mleft. #3 \mright|_{#4}}
   {#3#2|_{#4}}%
}

\usepackage{graphicx} 
\usepackage[colorinlistoftodos]{todonotes}
\usepackage{csquotes} %added because a warning told me to
\usepackage{upgreek}
\usepackage{ dsfont }
\usepackage{color}
\usepackage{textcomp}

\usepackage{float}
\usepackage{subfig}
\usepackage{mwe}

\usepackage{appendix}
\usepackage[a4paper,bindingoffset=0.2in,%
            left=1in,right=1in,top=1in,bottom=1in,%
            footskip=.25in]{geometry}
\usepackage[font=small, labelfont=bf, justification=centering]{caption}

\usepackage{lscape}
\usepackage{multirow}
\usepackage[natbib=true,citestyle=numeric-comp,bibstyle=numeric-comp,maxnames=1,sorting=none,maxbibnames=99]{biblatex}
\title{Understanding the effects of dichotomization of continuous outcomes on geostatistical inference}
\author[1]{Irene Kyomuhangi}
\author[2]{Tarekegn A. Abeku}
\author[2,3]{Matthew J. Kirby}
\author[2,4]{Gezahegn Tesfaye}
\author[1]{Emanuele Giorgi}
\affil[1]{CHICAS, Lancaster Medical School, Lancaster University, UK}
\affil[2]{Malaria Consortium, UK}
\affil[3]{London School of Hygiene and Tropical Medicine}
\affil[4]{PATH/MACEPA}

\date{}

\begin{document}
\maketitle

\begin{abstract}
Diagnosis is often based on the exceedance or not of continuous health indicators of a predefined cut-off value, so as to classify patients into positives and negatives for the disease under investigation. In this paper, we investigate the effects of dichotomization of spatially-referenced continuous outcome variables on geostatistical inference. Although this issue has been extensively studied in other fields, dichotomization is still a common practice in epidemiological studies. Furthermore, the effects of this practice in the context of prevalence mapping have not been fully understood. Here, we demonstrate how spatial correlation affects the loss of information due to dichotomization, how linear geostatistical models can be used to map disease prevalence and thus avoid dichotomization, and finally, how dichotomization affects our predictive inference on prevalence. To pursue these objectives, we develop a metric, based on the composite likelihood, which can be used to quantify the potential loss of information after dichotomization without requiring the fitting of Binomial geostatistical models. Through a simulation study and two applications on disease mapping in Africa, we show that, as thresholds used for dichotomization move further away from the mean of the underlying process, the performance of binomial geostatistical models deteriorates substantially. We also find that dichotomization can lead to the loss of fine scale features of disease prevalence and increased uncertainty in the parameter estimates, especially in the presence of a large noise to signal ratio. These findings strongly support the conclusions from previous studies that dichotomization should be always avoided whenever feasible.\\
\text{ }
\end{abstract}

\section{Introduction}
Continuous measurements of disease indicators - e.g. concentration of antibodies in a blood sample - are used in many branches of health research to aid diagnosis and treatment of patients, as well as monitoring and surveillance of diseases in populations. Diagnosis is often based on the exceedance or not of a cut-off value by the continuous indicator, to identify positives and negatives for the disease of interest \citep{coggon2009epidemiology}. In some cases, for instance in anaemia epidemiology, multiple cut-offs are also used to further categorize patients into groups, such as mild, moderate and severe \citep{world2011haemoglobin}. The rationale for such groupings is to aid and simplify both interpretation and presentation of the results \citep{del1997treatment,royston2006dichotomizing,bennette2012against}, while in clinical settings the groupings are used for targeted treatment. As a result, statistical analysis is often carried out on the categorical outcome obtained through the discretization of the continuous measurement.

The disadvantages and loss of information yielded by this practice have been investigated in several studies. \citet{fedorov2009consequences} showed that dichotomization of continuous outcome variables can lead to a loss of 36\% in terms of the Fisher's information for the population average. As a result of this, the statistical power required to estimate regression relationships between a health outcome and risk factors is also diminished \citep{altman2006cost}. Furthermore, in cases where the relationship is non-linear or non-monotonic, dichotomization or categorization into few classes may make that undetectable \citep{royston2006dichotomizing,bennette2012against}. All these issues are further exacerbated when the choice of specific cut-offs is inconsistent or, in some cases, even arbitrary \citep{Harrell2004,buettner1997problems,maccallum2002practice}. For example, cut-offs may vary within and across studies due to differences in the sample populations from which they are derived or due to changes in how they are defined according to clinical practice and operational policy. 

%\citet{austin2004inflation} showed that dichotomization can also increase the rate of false positive results in a logistic regression.

In this paper, we investigate the effects of dichotomization of spatially-referenced continuous outcome variables on geostatistical inference. Model-based geostatistics (MBG)  \citep{diggle1998mbg} is a likelihood-based paradigm that allows to carry out spatially continuous predictive inference on an outcome of interest using a spatially discrete set of data. Over the last two decades, MBG has been increasingly used to map the prevalence of infectious diseases \citep{diggle2016model}, especially in low-resource settings where disease registries are non-existent or geographically incomplete. In this context, there have been global efforts to increase the use of rapid diagnostic tests for diseases such as malaria and HIV \citep{incardona2017global,zhao2012adoption,kalla2019mass,world2015consolidated}, which are typically recorded as binary by labelling the tested individuals as either positive or negative. In other cases, instead, dichotomization is first carried out on a continuous disease indicator variable and a geostatistical model is then developed on the binary outcome. For example, in \citet{zimmerman2018}, a continuous score quantifying the deviation from normal growth in children is dichotomized in order map stunting prevalence; following a similar approach, \citet{magalhaes2011mapping} fit a binomial geostatistical model to dichotomized continuous haemoglobin densities so as to map anaemia prevalence in West Africa.

The effects of dichotomization on geostatistical inference have not been fully understood and, to the best of our knowledge, this is the first study that attempts to pursue this objective in the context of prevalence mapping. More specifically, in this paper, we provide answers to the following questions: 1) How does spatial correlation affect the loss of information and the estimation of regression relationships? 2) Can dichotomization lead to substantially different and more uncertain spatial predictions in disease prevalence? 3) How can we use linear geostatistical models to map disease prevalence and thus avoid the drawbacks of dichotomization? 

The structure of the paper is as follows. In section \ref{geostat_link} we describe the geostatistical modelling framework for disease prevalence mapping, and outline the differences and links between geostatistical models based on binary and continuous outcomes. In section \ref{sec:effects_dic}, we first explore the information loss due to dichotomization in terms of the Fisher's information for two observations. We then develop a metric which can be used to assess the loss of information for the estimation of the regression coefficients of any geostatistical model. We also carry out a simulation study to extend our investigation to the estimation of the covariance parameters and spatial predictions for prevalence. In section \ref{spat_pred} we illustrate two applications on the mapping of anaemia and stunting prevalence in Africa. Section \ref{sec:discussion} is a concluding discussion.

In what follows, fitting of geostatistical models and geostatistical prediction have been carried out using the Monte Carlo maximum likelihood method implemented in the \texttt{PrevMap} package \citep{giorgi2017prevmap} available from the Comprehensive R Network archive (\texttt{cran.r-project.org}).

\section{The link between geostatistical models for continuous and binary outcomes} \label{geostat_link}

Consider data from a cross-sectional survey where information on a continuous health outcome, $Y_{ij}$, is collected through examination of $n_i$ individuals residing at location $x_i$ for $j=1,\ldots,n_{i}$ and $i=1,\ldots,m$. We then assume that conditionally on a spatial Gaussian process $S=\{S(x):x\in\mathds{R}^2\}$, the $Y_{ij}$ are random Gaussian variables with mean $\mu_{ij}+S(x_i)$ and variance $\tau^2$. From the linear properties of Gaussian distributions, we can write the model in the following compact form

\begin{align} \label{eq:cont_spat}
Y_{ij} = \mu_{ij} + S(x_{i}) + Z_{ij},
\end{align}
where $Z_{ij}$ are i.i.d. Gaussian variables with mean zero and variance $\tau^2$, representing unexplained individual-level variation and the mean component $\mu_{ij}$ is modeled as a linear regression taking the form

$$
\mu_{ij} = \alpha+\beta^\top d(x_{i})+\gamma^\top e_{ij} 
$$
where we distinguish between covariates, $d(x_i)$, that express properties of the locations and covariates, $e_{ij}$, for individual traits e.g. age and gender.

We model $S(x)$ as a stationary and isotropic Gaussian process with mean zero, variance $\sigma^2$ and correlation function ${\rm Cor}\left\lbrace S(x),S(x^\prime)\right\rbrace = \rho \left( u \right)$ where $u=||x-x^\prime||$ denotes the Euclidean distance between $x$ and $x^\prime$. In the remainder of this paper, we shall define $\rho \left(\cdot\right)$ to be an exponentially decaying function with scale parameter $\phi$, i.e. $\rho(u) = \exp\{-u/\phi\}$.

Based on a predefined threshold $c$, whose exceedance or not defines the disease status of an individual, we define the binary outcome $\tilde{Y}_{ij}$ as
\begin{align}\label{eq:bin_spat}
\tilde{Y}_{ij}=
\begin{cases}
1 & \mbox{if } Y_{ij} < c \\
0 & \mbox{if }  Y_{ij} \geq c 
\end{cases}, 
\end{align}
with $\tilde{Y}_{ij}=1$ indicating a positive case for the disease under investigation and $\tilde{Y}_{ij}=0$ for a negative case.

From the model of the continuous outcome in \eqref{eq:cont_spat}, it follows that the model for $\tilde{Y}_{ij}$ is given by
\begin{eqnarray} \label{eq:probitmodel}
P\left(\tilde{Y}_{ij}=1 \middle | S(x_i)\right) &=& P\left(Y_{ij} < c  \middle | S(x_i) \right) \nonumber \\
&=& P\left(\frac{Y_{ij}-\mu_{ij}-S(x_i)}{\tau} < \frac{c-\mu_{ij}-S(x_i)}{\tau}  \middle | S(x_i)\right) \nonumber \\
&=& \Phi \left(\frac{c-\mu_{ij}-S(x_i)}{\tau} \right) = p_{ij}, 
\end{eqnarray} 
where $\Phi(\cdot)$ is the cumulative density function of a standard Gaussian variable.
Hence, the resulting model for $\tilde{Y}_{ij}$ is a Binomial geostatistical model with probit link function and linear predictor 
\begin{equation}
    \label{eq:lp_probit}
    \eta_{ij} = \Phi^{-1}(p_{ij}) = \tilde{\mu}_{ij}+\tilde{S}(x_i)
\end{equation}
where $\tilde{\mu}_{ij} = -\mu_{ij}/\tau$ and $\tilde{S}(x_i) = -S(x_i)/\tau$.

The functional relationships that link the parameters of the geostatistical model for $\tilde{Y}_{ij}$ with those of the model for $Y_{ij}$ are the following
\begin{equation}
    \label{eq:function_rel}
    \begin{cases}
    \tilde{\alpha}= (c-\alpha)/\tau \\
    \tilde{\beta} = -\beta/\tau \\
    \tilde{\gamma} = -\gamma/\tau \\
    \tilde{\sigma}^2 = \sigma^2/\tau^2 \\
    \end{cases}.
\end{equation}

The above equations can thus be used to obtain the parameter estimates for a geostatistical model for $\tilde{Y}_{ij}$ by transforming the parameter estimates from the geostatistical model for $Y_{ij}$. Note that it is not possible, instead, to map the estimates from the model for $\tilde{Y}_{ij}$ into those for $Y_{ij}$ because the parameter $\tau^2$ cannot be estimated from binary data. The unstructured component $Z_{ij}$ is in fact integrated out in \eqref{eq:lp_probit} and, as shown in \eqref{eq:function_rel}, all the parameters on the left hand-side are expressed as a ratio between $\tau$ and all other parameters in the model for $Y_{ij}$. Finally, $\phi$ is not included in \eqref{eq:function_rel}, since the scale of the spatial correlation of $\tilde{S}(x)$ is the same as that of $S(x)$.

\section{Quantifying the effects of dichotomization} \label{sec:effects_dic}

In Section \ref{subsec:unk_alphat}, we first study the loss of information due to dichotomization for the estimation of the mean of the process using an intercept-only model, when all other parameters are known. In Section \ref{subsec:simulation}, we carry out a simulation study to the more common case when all parameters are unknown. Here we study the effect on dichotomization on parameter estimation. 
In both sections, we shall restrict our attention to the scenario of a single observation per location, hence we set $n_{i}=1$ for all $i$ and drop the $j$ subscript. 

\subsection{Unknown regression coefficients and known covariance parameters}
\label{subsec:unk_alphat}

\subsubsection{Special case of m=2 for an intercept-only model}
The objective in this section is to quantify the loss of information in terms of the expected Fisher information (EFI) with respect to $\tilde{\alpha}$, the parameter which regulates the mean level of disease prevalence. Here, we restrict our attention to the simpler case of two observations at two locations, hence $m=2$ and $n_1=n_2=1$, for an intercept-only model. As it will be shown in the applications of Section \ref{spat_pred}, this simpler scenario provides useful insights on the effects of dichotmozation which are consistently observed in the case of more than two observations. A more general scenario, however, shall also be considered in the next section. 

We re-express the linear geostatistical model for a continuous outcome $Y_i$ as
\begin{align} \label{eq:cont_geostat_GLM}
Y_{i} = \alpha + S(x_i) + Z_{i},\text{ for }i=1,\ldots,m
\end{align}
where $S(x_i)$ is a stationary and isotropic Gaussian process with the same properties as defined in equation \eqref{eq:cont_spat}.

Let $\Sigma_{Y}=\Sigma+\tau^2I$ be the covariance matrix of the vector $Y=(Y_1,\ldots,Y_m)$, with $\Sigma$ and $I$ denoting the spatial covariance matrix with $(i,j)$-th entry $\sigma^2\exp\{-\|x_i-x_j\|/\phi\}$ and an $m$ by $m$ identity matrix, respectively. \par

In order to quantify the loss of information that arises from the dichotomization of the $Y_i$, we first re-parametrize the linear model in \eqref{eq:cont_geostat_GLM} with respect to the prevalence parameters as defined in \eqref{eq:function_rel}; note that $\alpha = c-\tau\tilde{ \alpha}$. We then obtain the EFI for $\tilde{\alpha}$ under the linear model, given by
\begin{align} \label{eq:FI_cont_spat}
I_Y(\tilde{\alpha})=\tau^2 \mathds{1}^T\Sigma_{Y}^{-1}\mathds{1},
\end{align} 
where $\mathds{1}$ is a vector with all entries equal to 1.

In the case of the dichotomized outcome $\tilde{Y}_{i}$, the computation of the EFI is further complicated by the fact that the log-likelihood function is not available in closed form. More specifically, this is given by the marginal distribution of the outcome $\tilde{Y}=(\tilde{Y}_1,\ldots,\tilde{Y}_{m})$, i.e.
\begin{equation}
\label{eq:lik_binary}
\log L(\tilde{\alpha}) = \log \left(\int_{\mathbb{R}^m} f(\tilde{s}) f(\tilde{y} | \tilde{s}; \tilde{\alpha}) \: d\tilde{s}\right),
\end{equation}
where $f(\tilde{s})$ is the density of a multivariate Gaussian distribution with mean zero and covariance matrix $\tilde{\Sigma} = \Sigma/\tau^2$, whilst
\begin{eqnarray}
\label{eq:y_given_s}
f(\tilde{y} | \tilde{s}; \tilde{\alpha}) &=& \prod_{i=1}^{m} f(\tilde{y}_{i} | \tilde{s}_{i}; \tilde{\alpha}) \nonumber \\
                   &=& \exp\left\{\sum_{i=1}^m\left[\tilde{y}_{i}\log\left(\frac{p_i}{1-p_i}\right)+(1-\tilde{y}_i)\log\{1-p_i\}\right]\right\} \nonumber \\
                   &=& \exp\{g(\tilde{y} | \tilde{s}; \tilde{\alpha})\}
\end{eqnarray}
where $\Phi^{-1}(p_i) = \tilde{\alpha} + \tilde{S}(x_i)$. To obtain the EFI for $\tilde{\alpha}$, we first take the second derivative of \eqref{eq:lik_binary} with respect to $\tilde{\alpha}$ to give
\begin{eqnarray}
\label{eq:2nd_der}
\frac{d^2 \log L(\tilde{\alpha})}{d^2 \tilde{\alpha}} &=& L^{-1}(\tilde{\alpha})\int_{\mathbb{R}^m} f(\tilde{s})f(\tilde{y} | \tilde{s}; \tilde{\alpha}) \bigg[\left(\frac{d g(\tilde{y} | \tilde{s}; \tilde{\alpha})}{d \tilde{\alpha}}\right)^2 + \nonumber \\
&& \frac{d^2 g(\tilde{y} | \tilde{s}; \tilde{\alpha})}{d^2 \tilde{\alpha}}\bigg]d \tilde{s}+\left(\frac{d \log L(\tilde{\alpha})}{d \tilde{\alpha}}\right)^2,
\end{eqnarray}
where  
$$
\frac{d \log L(\tilde{\alpha})}{d \tilde{\alpha}} = L^{-1}(\tilde{\alpha})\int_{\mathbb{R}^m} f(\tilde{s}) f(\tilde{y} | \tilde{s}; \tilde{\alpha}) \frac{d g(\tilde{y} | \tilde{s}; \tilde{\alpha})}{ d \tilde{\alpha}} \: d \tilde{s}.
$$
Finally, we average over the distribution of $\tilde{Y}$
$$
I_{\tilde{Y}}(\tilde{\alpha}) = E_{\tilde{Y}}\left[-\frac{d^2 \log L(\tilde{\alpha})}{d^2 \tilde{\alpha}}\right].
$$
Since the above quantity is not available in closed form we compute $I_{\tilde{Y}}(\tilde{\alpha})$ using Monte Carlo methods.

To quantify the loss of information, we then use the following metric $R(\tilde{\alpha}) = 1-I_{\tilde{Y}}(\tilde{\alpha})/I_{Y}(\tilde{\alpha})$. In the special case of $S(x)=0$ for all $x$, $R(\tilde{\alpha})$ reduces to
\begin{align} \label{eq:EFI_ratio_case1}
R(\tilde{\alpha})= 1-\left[\frac{[\Phi^{''}(\tilde{\alpha})][1-\Phi(\tilde{\alpha})]-[\Phi^{'}(\tilde{\alpha})]^2}{1-\Phi(\tilde{\alpha})}-\frac{[\Phi^{''}(\tilde{\alpha})][\Phi(\tilde{\alpha})]- [\Phi^\prime(\tilde{\alpha})]^2}{\Phi(\tilde{\alpha})}\right]
\end{align}
where $\Phi^{'}(\cdot)$ and $\Phi^{''}(\cdot)$ are the first and second derivative of $\Phi(\cdot)$, respectively. \citet{fedorov2009consequences} have shown that  $I_{\tilde{Y}}(\tilde{\alpha}) \leq I_{Y}(\alpha)$, and that the lower limit of \eqref{eq:EFI_ratio_case1} is about $36\%$. Also, note that \eqref{eq:EFI_ratio_case1} is not dependent on $m$.

\begin{figure} [ht]
\centering 
\includegraphics[width=1.05\textwidth]{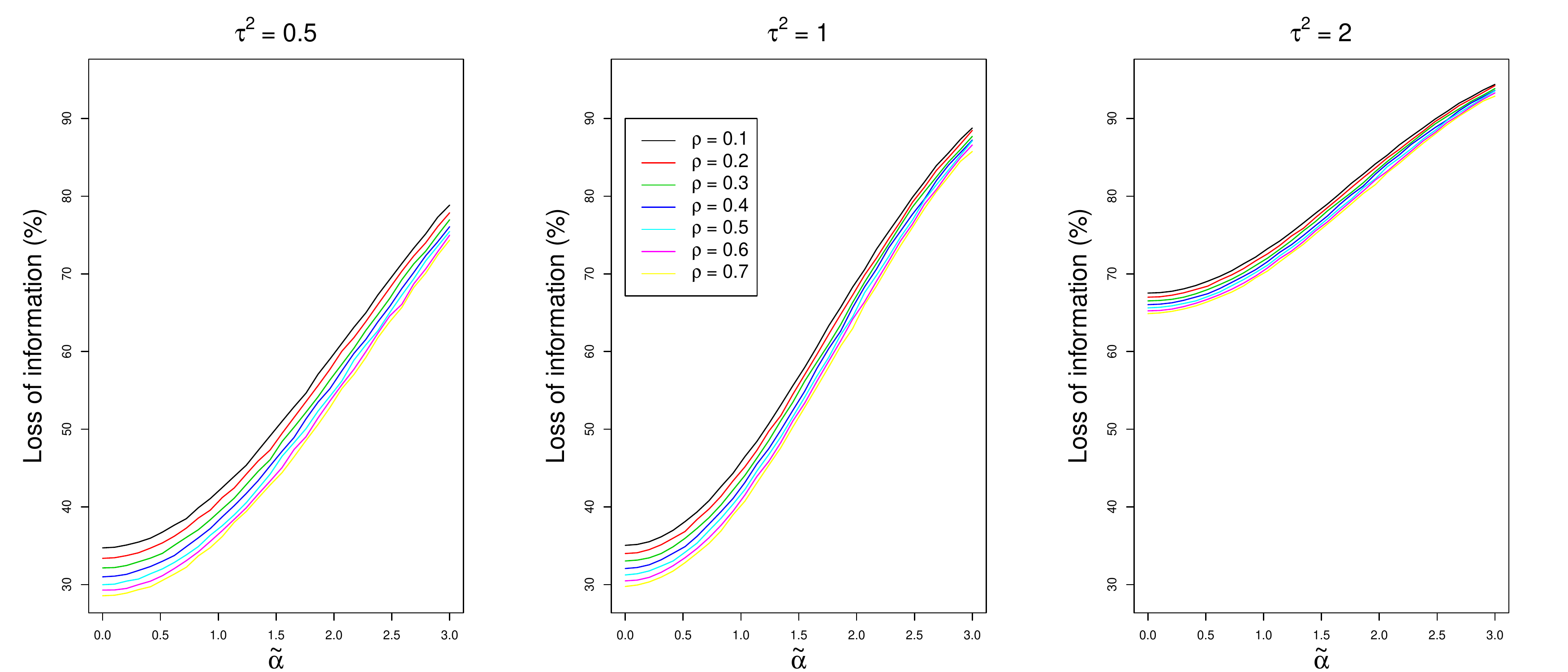}
\caption{Curves for $R(\tilde{\alpha})$, shown as a percentage, by fixing $\sigma^2=1$ and varying $\tau^2 \in \{0.5,1,2\}$ and the spatial correlation between two observations $\rho \in \{i/10; i=1,\ldots,7\}$.}
\label{fig:loss_info}
\end{figure}

To compute the integrals which define $-d^2 \log L(\tilde{\alpha})/d^2 \tilde{\alpha}$, we use a quadrature apporoach based on Quasi Monte Carlo methods. Finally, for the computation of the expectation in $I_{\tilde{Y}}(\tilde{\alpha})$, we use 10,000 samples and vary the spatial correlation between the two observations over $\rho\in\{i/10; i=1,\dots,7\}$. Figure \ref{fig:loss_info} shows different curves of $R(\tilde{\alpha})$, as a percentage, by setting $\sigma^2=1$ and letting $\tau^2$ vary over the set $\{0.5,1,2\}$. Notice that these curves are symmetric with respect to 0, although they are shown only for positive values of $\tilde{\alpha}$. Across all three panels of Figure \ref{fig:loss_info}, we observe that increasing values of $\rho$ lead to a reduction in the loss of information, although such reduction becomes smaller when the data are more noisy, i.e. when $\tau^2$ also increases. Most notably, the largest loss of information is observed for values of prevalence close to $100\%$ and $0\%$ corresponding to large positive and negative values for $\tilde{\alpha}$, respectively. The variance $\tau^2$ of the unstructured residuals $Z_i$ also plays a very important role as shown by the dramatic increase in $R(\tilde{\alpha})$ for $\tau^2=2$, with all curves placed above $R(\tilde{\alpha})=0.65$.

\subsubsection{General case m > 2 \label{subsec:cld}}

For the general case of more than two locations (i.e. $m>2$), the effects of dichotomization will also be dependent on the spatial arrangement of the sampled locations. In this section, we develop a metric that allows to quantify the potential loss of information due to dichotomization of continuous outcomes with respect the estimation of the regression coefficients $\theta=(\tilde{\alpha},\tilde{\beta},\tilde{\gamma})$ of any geostatistical model as defined by \eqref{eq:cont_spat}.

In order to lower the computational burden, we first approximate the likelihood function of both the continuous and dichotomized data using a composite likelihood approach \citep{varin2011}. More specifically, we consider all possible pairs of $\tilde{Y}_{ij}$ and treat each of these as independent bivariate distributions. Let $\tilde{Y}_h$ and $\tilde{Y}_k$ denote the vectors of binary outcomes associated with locations $x_h$ and $x_k$ and which are obtained through dichotomization of $Y_h$ and $Y_k$, respectively. We then write
\begin{equation}
\label{eq:composite_likelihood}
    L(\theta) \approx L_{CL}(\theta) = \prod_{h=1}^{m-1} \prod_{k=h+1}^m f\left(\tilde{y}_h, \tilde{y}_{k}; \tilde{\theta}\right).
\end{equation}
In the above equation the bivariate probability functions $f(\tilde{y}_h, \tilde{y}_{k}; \tilde{\theta})$ are expressed by the following integral in two dimensions

$$
f(\tilde{y}_h, \tilde{y}_{k}; \tilde{\theta}) = \int_{R^2} f(\tilde{s}) f\left(\tilde{y}_h, \tilde{y}_{k} | \tilde{s}; \tilde{\theta}\right) \: d \tilde{s}
$$
where $f\left(\tilde{y}_h, \tilde{y}_{k} | \tilde{s}; \tilde{\theta}\right)$ consists of a product of $n_h+n_k$ probability functions for the binary observations in $y_h$ and $y_k$. \par

Let $\hat{\theta}_{LM}$ denote the maximum likelihood estimates of $\theta$ obtained from the linear model using the system of equations in \eqref{eq:function_rel}. In order to understand how more or less dispersed the composite likelihood becomes after dichotmization, we proceed as follows. We first compute the second derivative of the composite log-likelihood at $\hat{\theta}_{LM}$, i.e.
$$
H_{\tilde{Y}}(\hat{\theta}_{LM}) = \left[\frac{\partial^2 \log L_{CL}}{\partial \theta_i \partial \theta_j}\right]_{\theta=\hat{\theta}_{LM}}.
$$
For a continuous outcome $Y$, we have
$$
H_{Y}(\theta) = -\tau^2 \sum_{h=1}^{m-1}\sum_{k=h+1}^{m} D_{h}^\top \Sigma_{hk}^{-1} D_{k}
$$
where $\Sigma_{hk}$ is covariance matrix between $Y_h$ and $Y_k$, and $D_{h}$ and $D_{k}$ are the design matrices associated with locations $x_h$ and $x_k$, respectively. To quantify the change in the dispersion of the composite likelihood around $\hat{\theta}_{LM}$, we finally compute
\begin{eqnarray*}
CLD(\hat{\theta}_{LM}) &=& \log\{{\rm det}[-H^{-1}_{\tilde{Y}}(\hat{\theta}_{LM})]\}- \log\{{\rm det}[-H^{-1}_{Y}(\hat{\theta}_{LM})]\} \\
&=& \log\{{\rm det}[-H_{Y}(\hat{\theta}_{LM})]\}- \log\{{\rm det}[-H_{\tilde{Y}}(\hat{\theta}_{LM})]\}
\end{eqnarray*}
where ${\rm det}(A)$ denotes the determinant of a square matrix $A$. Large values of $CLD(\hat{\theta}_{LM})$ indicate a more dispersed composite likelihood around $\hat{\theta}_{LM}$ for the binary data $\tilde{Y}$, which we interpret as evidence of loss of information due to dichotmization. Also, note that computation of CLD can be carried out without fitting any Binomial geostatatistical model. 

In the applications of Sections \ref{subsec:anaemia} and \ref{subsec:stunting}, we compute the CLD metric, using by plugging-in the maximum likelihood estimates for the covariance parameters from the linear geostatistical models. 

\subsection{Simulation study}
\label{subsec:simulation}
The objective of this simulation is to quantify the effects of dichotomization on parameter estimation and spatial predictions of prevalence. To this end we consider the linear model for $Y_i$ as specified in \eqref{eq:cont_geostat_GLM} and its dichotomization using a threshold $c$ to give $\tilde{Y}_i=1$ if $Y_i < c$ and 0 otherwise.

In the simulation, we set $\alpha=0$ and $\sigma^2=1$. We consider several scenarios obtained through all possible combinations of values for the model parameters defined in Table \ref{table:SimulationConditions}. For a given scenario, we simulate 1,000 data-sets of both continuous and dichotomized outcomes and fit their corresponding geostatistical models. We use a regular grid covering the unit square with spacing of 1/14 to give a sample size of $m=225$. For each of the fitted models, we extract the estimates for $\tilde{\alpha}$, $\tilde{\sigma}^2$ and $\phi$, and predict prevalence $p_{i} = \Phi\left(\tilde{\alpha}+\tilde{S}(x_i)\right)$ at each of the grid points. We summarise the results using the bias and mean square error (MSE) and, for the prevalence predictions, average these two indices over the grid locations.

Tables \ref{tab:param_sim_results} and \ref{tab:pred_sim} report the results for the model parameters and spatial predictions for prevalence, respectively. Overall, bias and MSE for $\tilde{\alpha}$ and $\phi$ are consistently smaller in the model fitted to the continuous data (C) than for that fitted to the binary data (B). In the case of $\tilde{\sigma}^2$, instead, the performance of both models is strongly affected by the scale of the spatial correlation $\phi$: for $\phi=0.1$ the model B outperforms model C in terms of bias and MSE, whilst the opposite is observed for $\phi=0.2$. A possible explanation for this may be due to the fact that in the linear geostatistical model, higher spatial correlation helps to better separate the individual contributions of the signal component $S(x_i)$ and the noise component $Z_i$ to the total variation in the outcome $Y_i$, thus improving the estimation of $\tilde{\sigma}^2=\sigma^2/\tau^2$. In the case of the binary data, instead, $\tilde{S}(x_i)$ is the only source of over-dispersion and, as a result of this, a higher spatial correlation leads to a larger number of concordant binary outcomes and, therefore, to a poorer estimate of the variance of $\tilde{S}(x_i)$. Also, we notice that the estimation of $\tilde{\sigma}^2$ and $\phi$ does not appear to be affected by the threshold $c$, unlike $\tilde{\alpha}$. Finally, the results for the spatial predictions of prevalence show that the performance of model C is unaffected by changes in $c$ and $\tau^2$, while for $\phi=0.2$ the predictions have slightly lower MSE than for $\phi=0.1$. Model B, instead, delivers predictions with higher bias for increasing $c$ which can be partly explained by the positive increase in the bias in the estimates of $\tilde{\alpha}$ for increasing $c$.

We have also conducted further simulations under the same scenarios defined in Table \ref{table:SimulationConditions} but for a larger sample size $m=450$, by placing additional points on a regular grid adjacent to the unit square so as to cover the rectangle $[0,2]\times[0,1]$. The results, reported in the Appendix (Tables \ref{tab:pred_sim_n450} and \ref{tab:pred_sim_n450}), lead to the same conclusions drawn for $m=225$.

\begin{table}[ht]
\centering
\caption{True values for $\tau^2$, $\phi$ and $c$ used in the simulation study.}
\begin{tabular}{p{6cm} p{2.5cm} p{3cm}} \hline
 & Symbol & Variations \\
\hline
Variance of the nugget effect & $\tau^2$ & 0.5, 1, 2 \\
True scale of spatial correlation & $\phi$ & 0.1, 0.2 \\
Cut-off & c & 0, 0.2, 0.4 \\
\hline 
\end{tabular}
\label{table:SimulationConditions}
\end{table}

\begin{table}[ht]
\centering
\caption{Bias and mean square error (in brackets) for $\tilde{\alpha}$, $\tilde{\sigma}^2$ and the estimate $\hat{\phi}$ obtained from the geostatistical models fitted the binary (B) and continuous (C) outcomes. \label{tab:param_sim_results}}
\resizebox{\columnwidth}{!}{%
\begin{tabular}{ccc|cc|cc|cc} \hline
 & & & \multicolumn{2}{|c|}{c=0} & \multicolumn{2}{|c|}{c=0.2} & \multicolumn{2}{|c}{c=0.4} \\
Parameter & $\tau^2$ & $\phi$ & B & C & B & C & B & C \\
\hline
$\tilde{\alpha}$ & 0.5 & 0.1 &  0.009 (0.168) & 0.005 (0.113) & 0.063 (0.312) & 0.017 (0.115) & 0.157 (0.263) & 0.042 (0.139) \\
& 1 & 0.1 & -0.009 (0.126) & -0.008 (0.068) & 0.060 (0.409) & 0.004 (0.064) & 0.153 (0.119) & 0.041 (0.080) \\
& 2 & 0.1 & -0.006 (0.079) & -0.003 (0.036) & 0.077 (0.493) & 0.022 (0.040) & 0.148 (0.169) & 0.050 (0.054) \\ 
& 0.5 & 0.2 & -0.025 (0.624) & -0.013 (0.296) & 0.156 (0.648) & 0.048 (0.282) & 0.238 (0.722) & 0.028 (0.303) \\
& 1 & 0.2 & -0.018 (0.332) & -0.007 (0.150) & 0.093 (0.585) & 0.008 (0.137) & 0.215 (0.323) & 0.025 (0.145) \\
& 2 & 0.2 & -0.007 (0.185) & -0.004 (0.080) & 0.106 (0.566) & 0.023 (0.080) & 0.164 (0.272) & 0.022 (0.088) \\ \hline
$\tilde{\sigma}^2$& 0.5 & 0.1 & -0.011 (1.296) & 0.788 (5.843) & -0.033 (1.324) & 0.604 (5.170) & 0.022 (1.594) & 0.734 (5.741) \\
& 1 & 0.1 & 0.234 (0.787) & 0.822 (5.953) & 0.190 (0.688) & 0.750 (5.255) & 0.224 (0.701) & 0.741 (5.942) \\
& 2 & 0.1 & 0.211 (0.326) & 0.672 (5.047) & 0.217 (0.364) & 0.600 (4.961) & 0.204 (0.324) & 0.653 (4.833) \\ 
& 0.5 & 0.2 & 1.641 (8.894) & 0.527 (3.091) & 1.784 (12.672) & 0.574 (3.268) & 1.566 (8.464) & 0.515 (3.401) \\
& 1 & 0.2 & 1.162 (3.755) & 0.399 (1.712) & 1.064 (3.243) & 0.372 (1.619) & 1.048 (2.949) & 0.365 (1.671) \\
& 2 & 0.2 & 0.548 (0.871) & 0.254 (1.104) & 0.575 (1.074) & 0.304 (1.829) & 0.534 (0.910) & 0.341 (1.707) \\ \hline
$\hat{\phi}$ & 0.5 & 0.1 & 0.088 (0.016) & 0.007 (0.002) & 0.084 (0.015) & 0.009 (0.002) & 0.085 (0.015) & 0.007 (0.002) \\
& 1 & 0.1 & 0.071 (0.014) & 0.004 (0.003) & 0.072 (0.014) & 0.006 (0.004) & 0.074 (0.015) & 0.006 (0.003) \\
& 2 & 0.1 & 0.060 (0.017) & 0.010 (0.007) & 0.063 (0.022) & 0.009 (0.007) &  0.056 (0.014) & 0.010 (0.006) \\ 
& 0.5 & 0.2 & 0.076 (0.029) & -0.017 (0.011) & 0.083 (0.034) & -0.020 (0.008) & 0.076 (0.030) & -0.021 (0.009) \\
& 1 & 0.2 & 0.068 (0.030) & -0.014 (0.016) & 0.058 (0.027) & -0.028 (0.010) & 0.061 (0.034) & -0.023 (0.013) \\
& 2 & 0.2 & 0.032 (0.025) & -0.024 (0.019) & 0.037 (0.031) & -0.024 (0.016) & 0.028 (0.019) & -0.032 (0.015) \\ 
\hline
\end{tabular}
}
\end{table}

\begin{table}[ht]
\centering
\caption{Bias and mean square error (in brackets), averaged over a 1/14 by 14 regular grid covering the unit square (hence, $m=225$), for the spatial predictions of prevalence obtained from the geostatistical models fitted to the binary (B) and continuous (C) outcomes. \label{tab:pred_sim}}
\resizebox{\columnwidth}{!}{%
\begin{tabular}{cc|cc|cc|cc} \hline
 & & \multicolumn{2}{|c|}{c=0} & \multicolumn{2}{|c|}{c=0.2} & \multicolumn{2}{|c}{c=0.4} \\
 $\tau^2$ & $\phi$ & B & C & B & C & B & C \\
\hline
0.5 & 0.1 & 0.001 (0.060) & 0.001 (0.039) & 0.018 (0.059) & 0.001 (0.038) & 0.034 (0.058) & 0.001 (0.036) \\
1 & 0.1 & -0.001 (0.051) & 0.001 (0.038) & 0.018 (0.051) & 0.001 (0.038) & 0.038 (0.050) & 0.001 (0.036) \\
2 & 0.1 & -0.001 (0.040) & 0.001 (0.033) & 0.020 (0.040) & -0.001 (0.033) & 0.038 (0.040) & -0.002 (0.032)\\
0.5 & 0.2 & -0.001 (0.042) & 0.001 (0.030) & 0.013 (0.042) & -0.001 (0.030) & 0.025 (0.041) & -0.001 (0.029) \\
1 & 0.2 & 0.001 (0.037) & 0.001 (0.028) & 0.014 (0.037) & -0.001 (0.028) & 0.030 (0.036) & -0.001 (0.027) \\
2 & 0.2 & -0.001 (0.031) & 0.001 (0.024) & 0.019 (0.031) & -0.001 (0.024) & 0.034 (0.031) & -0.002 (0.024) \\ \hline
\end{tabular}
}
\end{table}

\section{Applications} \label{spat_pred}

\subsection{Mapping anaemia prevalence in Ethiopia}
\label{subsec:anaemia}

In this section, we analyse data collected from the Beyond Garki project \footnote{\texttt{www.malariaconsortium.org/beyondgarki}}. This project consisted of cross-sectional surveys which were conducted in selected study sites in Ethiopia and Uganda to monitor changes in malaria risk in the context of interventions that had been implemented. The study sites were defined as a \lq health centre and the catchment population in selected villages around it\rq. Here we subset the data for the Hembecho site, in Ethiopia, collected during the 2012 survey, where a random sample of households were selected from a list of enumerated households from all villages within a radius of 2 to 6 kilometers of the health facility. Among the data obtained in this survey were continuous measurements of haemoglobin density (g/dL), taken from blood samples of individuals living in the households. These measurements were then used to determine the anaemia status of individuals. Further details of the study design and data collection can be found in  \citet{abeku2015monitoring}.

In this analysis, the objective is to identify areas where the anaemia prevalence is highly likely to exceed a 20$\%$ threshold for 20 year old women. Hence, we map and compare exceedance probabilities from the geostatistical models for the continuous and binary outcomes. The chosen threshold for anaemia prevalence has clinical, operational and public health significance for policy decisions, with the World Health Organisation (WHO) classifying 20\% anaemia prevalence as \lq moderate public health significance\rq \citep{world2011haemoglobin}. Finally, the rationale for carrying out predictions for 20 years old women is that one of the key WHO Global Nutrition Targets for 2025 is a 50\% reduction of anaemia in women of reproductive age \citep{world2011haemoglobin,world2014globalpol}, which is defined as 15-49 years \citep{WRA}.

\begin{figure} [ht]
\centering %centers figure
\includegraphics[width=1\textwidth]{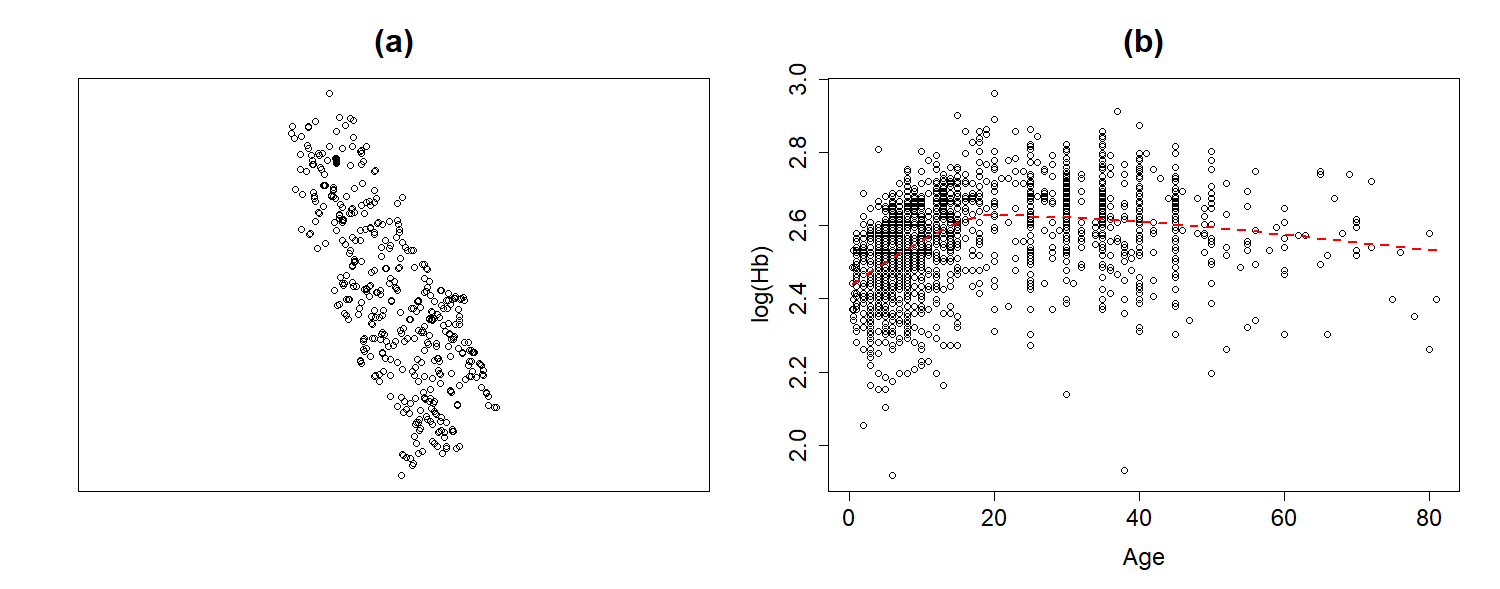}
\caption{(a) locations of the households in the survey; (b) scatter plot of the log-transformed haemoglobin density against age, in years. The dashed red line in the in panel (b) is a least square fit of the linear spline defined in the main text of Section \ref{subsec:anaemia}.}
\label{fig:/hembecho_covars}
\end{figure}

The data-set contains information on 1712 individuals distributed over 457 households, with an average of 3.7 individual in each household. The continuous outcome variable, $Y_{ij}$, is the log-transformed haemoglobin density for the $j$-th individual at the $i$-th household. To account for the non-linear relationship between the log-transformed anaemia density and age, as shown in Figure \ref{fig:/hembecho_covars}, we use a linear spline with knots at 15 and 30 years. Our proposed linear model for $Y_{ij}$ is thus expressed as
\begin{align}
Y_{ij} = \alpha + \sum_{h=1}^{3}\beta_h b_h(a_{ij}) + \beta_4 d_{ij} + S(x_i) + Z_{ij},
\end{align}
where: $a_{ij}$ is the age, in years, of an individual; $d_{ij}$ is a binary indicator of the sex of an individual, with ``female'' as reference category; 
$b_{h}(\cdot)$ are the base functions of the linear spline defined as $b_1(a) = a$, $b_2(a) = max\{0,a-15\}$ and $b_3(a) = max\{0,a-30\}$.

Dichotomization of $Y_{ij}$ results in the binary  outcome variable $Y^*_{ij}$ indicating anaemia status, where $Y^*_{ij}=1$ denoting a positive case and $Y^*_{ij}=0$ a negative case for severe anaemia. In order to classify an individual as positives or negatives, thresholds for severe anaemia have been applied using individual-level information on age, sex and pregnancy status as defined in Table \ref{tab:anaemia cutoffs}. As result of this, we then modify equation \eqref{eq:probitmodel} as
$$
p_{ij} = \Phi\left(\frac{c_{ij}-\mu_{ij}-S(x_i)}{\tau}\right)
$$
where $c_{ij}$ is the logarithm of the threshold values which must now be incorporated as an additional covariate into the linear predictor, i.e.
$$
\Phi^{-1}(p_{ij}) = \tilde{\alpha} + \sum_{h=1}^{3}\tilde{\beta}_h b_h(a_{ij}) + \tilde{\beta}_4 d_{ij} + \tilde{\beta}_{5}c_{ij}+\tilde{S}(x_i)
$$
where $\tilde{\beta}_5=1/\tau$.

\begin{table}[ht]
\small
\centering
\caption{Thresholds of haemoglobin densities (g/dL) for anaemia diagnosis \citep{world2011haemoglobin}.}
\label{tab:anaemia cutoffs}
\begin{tabular}{lcccccccr} \hline
& &Anaemia& \\
Age or Sex group     & Mild & Moderate & Severe  \\ \hline
Children (Age 6-59 months)         & 10.0-10.9     & 7.0-9.9   & $<7.0$  \\
Children (Age 5-11 yrs)         & 11.0-11.4     & 8.0-10.9   & $<8.0$  \\
Children (Age 12-15yrs)         & 11.0-11.9     & 8.0-10.9   & $<8.0$  \\
Pregnant women (Age > 15yrs)          & 10.0-10.9      & 7.0-9.9   & $<7.0$  \\
Non-pregnant women (Age > 15 yrs)         & 11.0-11.9      & 8.0-10.9   & $<8.0$  \\
Men (Age > 15 yrs)         & 11.0-12.9      & 8.0-10.9   & $<8.0$  \\
\hline
\end{tabular}
\end{table} 

Table \ref{tab:anaemia_estim} reports the maximum likelihood estimates and 95\% confidence intervals of the model parameters for the binary and continuous outcomes. The linear geostatistical model gives an estimate for $\tau^2$, the variance of $Z_{ij}$, of about $1.133\times 10^{-2}$ (95$\%$ CI: $1.050\times 10^{-2}$, $1.222\times 10^{-2}$)  and for $\sigma^2$ of about $1.558\times 10^{-3}$ (95$\%$ CI: $0.954\times 10^{-3}$, $2.422\times 10^{-3}$), yielding an estimated noise to signal ratio $\tau^2/\sigma^2$ of about $7.3$. The CLD metric (Section \ref{subsec:cld} ) is 771.235 which indicates a larger dispersion of the composite likelihood for the binary data than for the continuous data.  We observe that the estimates of the parameters are all comparable except for $\tilde{\sigma}^2$ ($= \sigma^2/\tau^2$), as indicated by the non-overlapping confidence intervals from the two models. More importantly, we observe that the confidence intervals for the regressions coefficients are narrower for the linear model.

\begin{figure}[ht]
\centering
\includegraphics[width=1\linewidth]{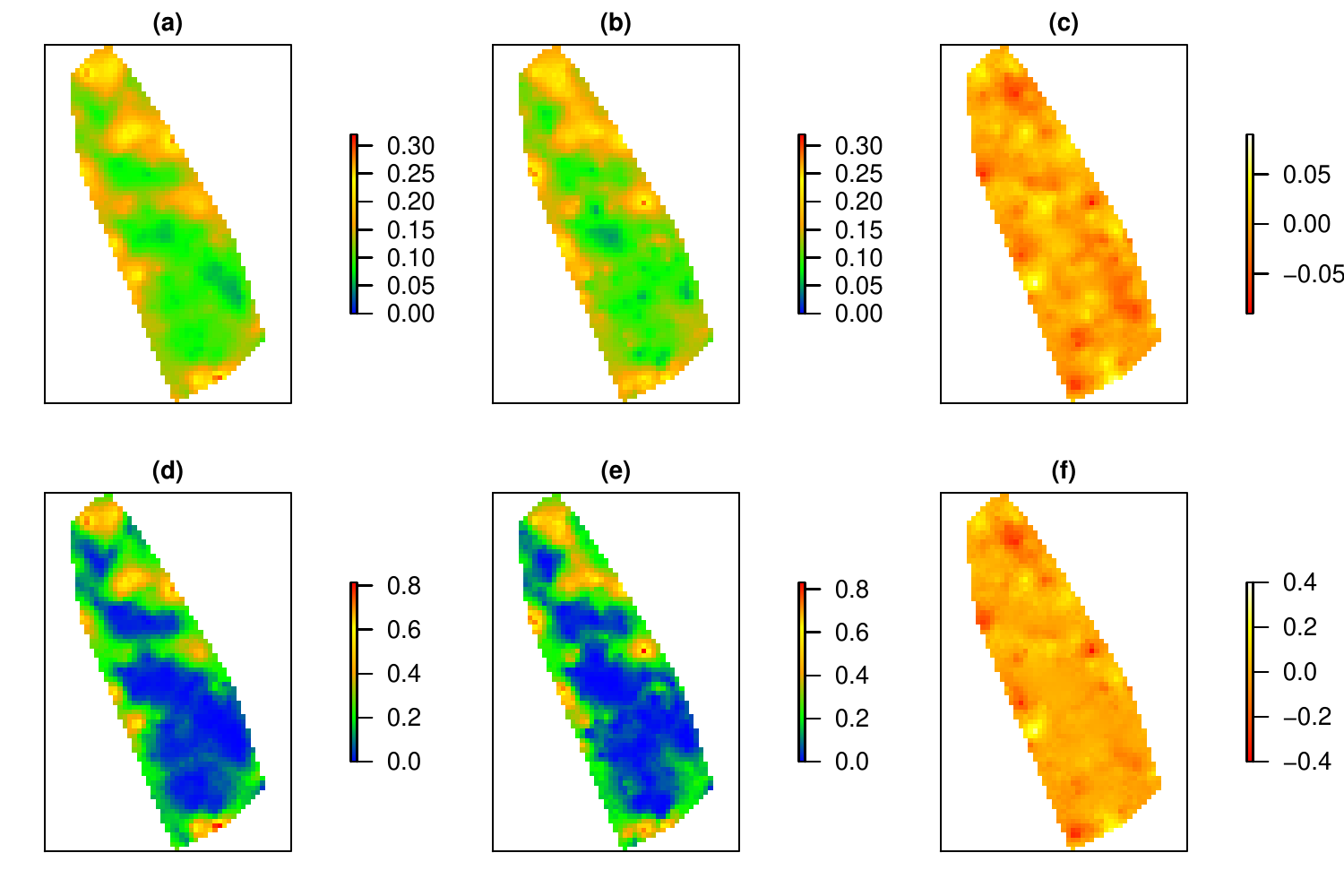}
\caption{Predicted anaemia prevalence for a 20 year old woman. Upper panels: prevalence surfaces from the binomial model (a) and the continuous models (b), and the difference between the first and the second (c). Lower panels: exceedance probabilities for a 20$\%$ prevalence threshold obtained from the binary (d) and the continuous models (e), and their difference (f).}
\label{fig:anem_pred}
\end{figure}

Figure \ref{fig:anem_pred} shows the resulting  anaemia prevalence predictions for 20 year old women from the two models.  While the overall pattern of predicted prevalence is similar between the models (see Figures \ref{fig:anem_pred}(a) and \ref{fig:anem_pred}(b)), there are non-negligible differences ranging from -8.23\% to 7.85\% prevalence (Figure \ref{fig:anem_pred}(c)). Similarly, the maps of the exceedance probability qualitatively show similar spatial patterns (figures \ref{fig:anem_pred}(d) and \ref{fig:anem_pred}(e)). However, we identify small areas where the differences range from -38.6\% to 31.20\% (Figure \ref{fig:anem_pred}(f)). 

\begin{table}[ht]
\centering
\caption{Maximum likelihood estimates with associated $95\%$ confidence intervals (CI) for the geostatistical models fitted to the anaemia data. \label{tab:anaemia_estim}}
\begin{tabular}{lrcrc}
  \hline
     & \multicolumn{2}{c}{Binomial model} & \multicolumn{2}{c}{Linear model} \\
Term & Estimate & 95$\%$ CI & Estimate & 95$\%$ CI \\
  \hline
$\tilde{\beta}_1$ & 0.079 & (-0.220, 0.378) & -0.278 & (-0.377, -0.179) \\ 
$\tilde{\beta}_2$ & -0.066 & (-0.124, -0.008) & -0.115 & (-0.131, -0.100) \\ 
$\tilde{\beta}_3$ & 0.052 & (-0.025, 0.129) & 0.097 & (0.072, 0.122) \\ 
$\tilde{\beta}_4$ & 0.071 & (0.021, 0.121) & 0.050 & (0.031, 0.069) \\ 
$\tilde{\sigma}^2$ & 0.527 & (0.395,0.705) & 0.138 & (0.082, 0.218) \\ 
  $\phi$ & 0.325 & (0.201,0.396) & 0.250 & (0.093, 0.549) \\ 
   \hline
\end{tabular}
\end{table}

\subsection{Mapping stunting prevalence in Ghana}
\label{subsec:stunting}
The data analysed in this section are from the 2014 Demographic and Health Survey\footnote{\texttt{dhsprogram.com}} (DHS) conducted in Ghana. DHS are nationally representative household surveys conducted about every 5 years, and provide data on health and population indicators for monitoring and impact evaluation across Africa. The DHS surveys follow a stratified two-stage cluster design where in the first stage, enumeration areas are selected from previous population census files, followed by a second stage where, for each selected enumeration area, samples of households are sampled from updated lists of households to generate the so called sampling clusters. The GPS locations of a sampling cluster is then assigned to each of the individuals falling within that cluster.

Among the health indicators collected in this survey are anthropometric measurements, which are used to calculate the height-for-age Z-score (HAZ). HAZ are standardized scores which indicate the standard deviation from the mean of children's heights based on the WHO growth standards \citep{world2006child,who2006child} and are comparable across ages and sex. HAZ values below $-2$ are taken as an indication of stunted growth. 

One of the key WHO Global Nutrition Targets for 2025 is a 40\% reduction in the number of children under-5 who are stunted \citep{world2014globalstunt,world2014globalpol}. Additionally, a stunting prevalence above $40\%$ is considered  a high public health significance \citep{world2010nutrition}. Accordingly, we aim to map the exceedance probability of 40$\%$ stunting prevalence for a 2 year old who falls in the lowest wealth index category and whose mother has poor education. 

\begin{figure}[ht]
\centering %centers figure
\includegraphics[width=1\textwidth]{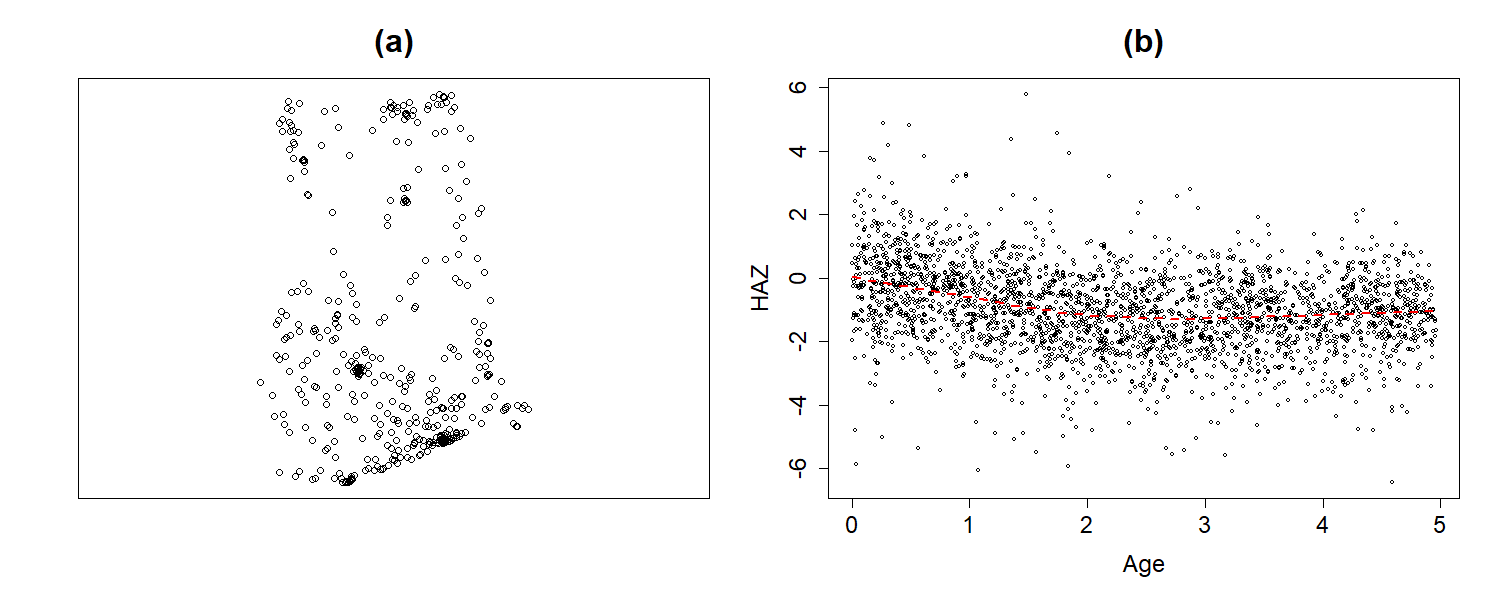}
\caption{Figure (a) shows the spatial distribution of households included in the analysis, while (b) shows the relationship between HAZ and age. The red dashed line in panel (b) corresponds to a least square fit of the linear spline defined in \eqref{eq:stunting_cont_model}.}
\label{fig:/ghana_covars}
\end{figure}

The data include information on children under 5 years old, with a total of 2671 sampled children and 410 clusters, giving an average of 6.5 children per cluster. The continuous outcome variable, $Y_{ij}$, is the HAZ for child $j$ in cluster $i$. Figure \ref{fig:/ghana_covars} (a) shows the empirical relationship between HAZ and age in years. Using a similar approach of the previous analysis, we capture the non-linear relationship with a linear spline having knots at 1 and 2 years. Hence, the resulting linear geostatistical model is  
\begin{equation}
\label{eq:stunting_cont_model}
Y_{ij} = \alpha + \sum_{h=1}^{3}\beta_h b_h(a_{ij}) + \beta_4 d_{ij} +\beta_5 e_{ij} + S(x_i) + Z_{ij},
\end{equation}
where: $a_{ij}$ is the age of a child; the basis functions of the linear splines are $b_{1}(a)=a$, $b_{2}(a)=\max\{0,a-1\}$ and $b_{2}(a)=\max\{0,a-2\}$; $d_{ij}$ is a score of maternal education, taking integer values from 1="Poorly educated" to 3="Highly educated"; $e_{ij}$ is a wealth index of the household, taking integer values from 1="Poor" to 3="Rich".

The maximum likelihood estimates and associated 95$\%$ confidence intervals are shown in Table \ref{tab:stunting_estim}. We also report that the estimates for $\sigma^2$ and $\tau^2$ from the linear geostatistical model are 0.071 (95$\%$ CI: 0.037, 0.126) and 1.396 (95$\%$ CI: 1.318, 1.477), respectively. Hence, the estimated ratio $\tau^2/\sigma^2$ is about 20, indicating that the data are substantially more noisy than those analysed in the previous section. This is also reflected in the CLD metric yielding a value of 9667.012 which is substantially larger than that reported for the anaemia analysis. Following from the results of Section \ref{sec:effects_dic}, this suggests that the effects of dichotomization on geostatistical inference  will also be stronger. We observe that the estimates of the regression coefficients are concordant in sign but the size of the effects of the covariates are different as indicated by the non-overlapping confidence intervals; as in the previous section, we observe that the confidence intervals for the regression coefficients from the linear model are all narrower. The estimated $\tilde{\sigma}^2$ and $\phi$ are also substantially different, with the linear geostatistical model providing lower estimates and narrower confidence intervals for both parameters.

\begin{figure} [ht]
\centering
\includegraphics[width=1\linewidth]{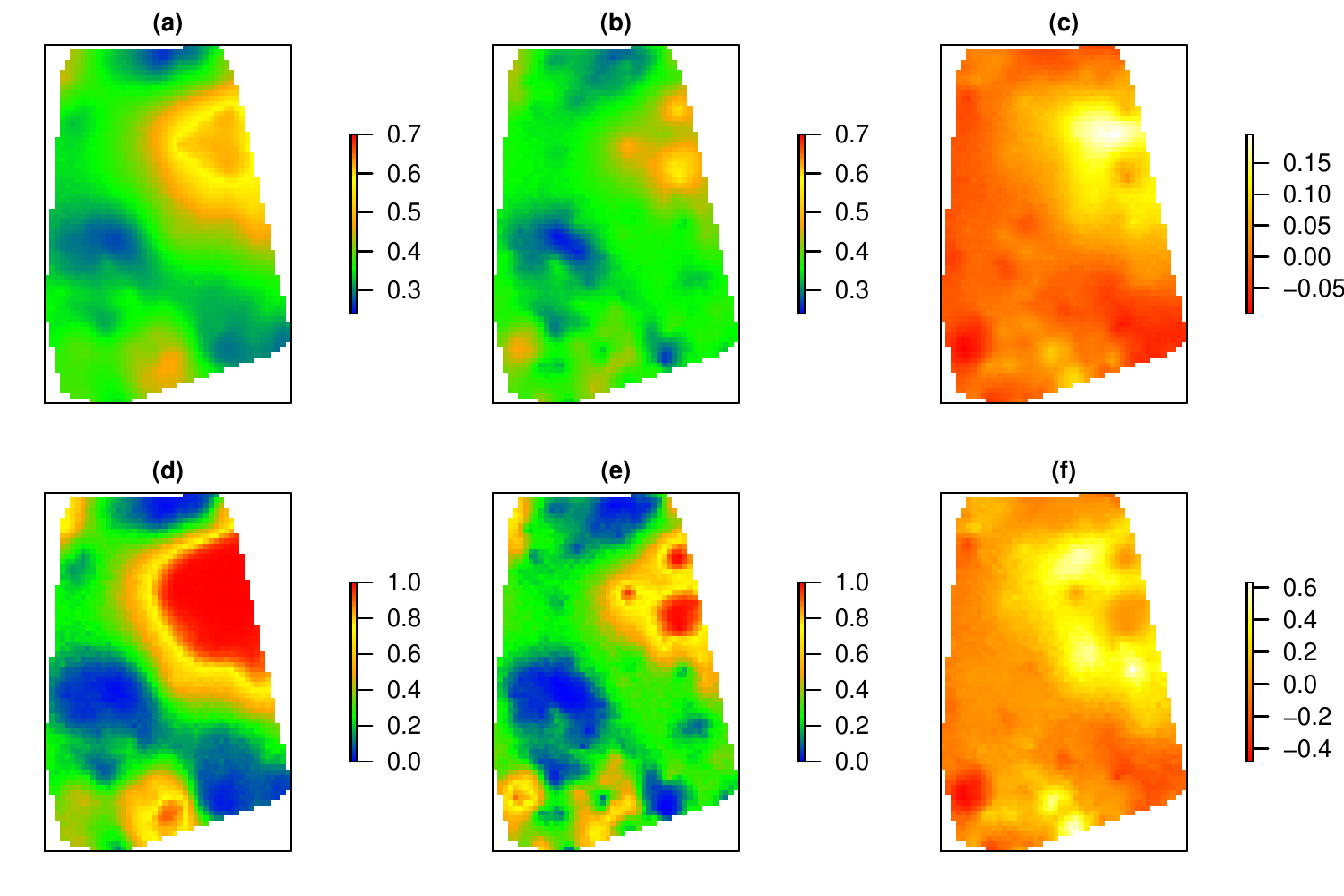} 
\caption{Predicted stunting prevalence for a 2 year old who falls in the lowest wealth index category and whose mother has poor education. Upper panels: prevalence surfaces from the binomial model (a) and the continuous models (b), and the difference between the first and the second (c). Lower panels: exceedance probabilities for a 40$\%$ prevalence threshold obtained from the binary (d) and the continuous models (e), and their difference (f).}
\label{fig:st_pred}
\end{figure}

The differences in the parameter estimates are also reflected in Figure \ref{fig:st_pred} which shows the predicted surfaces of stunting prevalence and the exceedance probabilities  from the two models. These predictions are for a  for a 2 year old who falls in the lowest wealth index category, and whose mother has poor education. Qualitatively, both models identify high and low levels of prevalence in the same areas. However, the differences in the predicted prevalence between the binomial model (Figure \ref{fig:st_pred}(a)) and the continuous model (Figure \ref{fig:st_pred}(b)), range from -9.38\% to 19.98\% prevalence (Figure \ref{fig:st_pred}(c)). Most notably, the binomial model presents much smoother maps than those from the linear model. For example, the binomial model identifies a single large hot-pot in the eastern part of Ghana, as being highly likely to exceed 40$\%$. The linear model, instead, shows three neighbouring hot-spots in the same area. The differences in exceedance probabilities between the two models range from -51.50\% to 64.90\% (Figure \ref{fig:st_pred}(f))

\begin{table}[ht]
\centering
\caption{Maximum likelihood estimates with associated $95\%$ confidence intervals (CI) for the geostatistical models fitted to the data on childhood malnutrition. \label{tab:stunting_estim}}
\begin{tabular}{lrcrc}
  \hline
     & \multicolumn{2}{c}{Binomial model} & \multicolumn{2}{c}{Linear model} \\
Term & Estimate & 95$\%$ CI & Estimate & 95$\%$ CI \\
  \hline
$\tilde{\beta}_{1}$ & 0.772 & (-0.002, 1.545) & 0.554 & (0.338, 0.771) \\ 
$\tilde{\beta}_{2}$ & 0.774 & (-0.331, 1.879) & 0.168 & (-0.168, 0.502) \\ 
$\tilde{\beta}_{3}$ & -1.942 & (-2.463, -1.421) & -0.830 & (-1.024, -0.639) \\ 
$\tilde{\beta}_{4}$ & -0.721 & (-1.174, -0.269) & -0.139 & (-0.239, -0.039) \\ 
$\tilde{\beta}_{5}$ & -0.444 & (-0.666, -0.221) & -0.259 & (-0.332, -0.186) \\ 
  $\tilde{\sigma}^2$ & 0.256 & (0.102, 0.528) & 0.051 & (0.026, 0.091) \\ 
  $\phi$ & 157.301 & (59.706, 341.451) & 51.899 & (14.657, 136.232) \\ 
   \hline
\end{tabular}
\end{table}

\section{Discussion}
\label{sec:discussion}
Understanding of the effects of dichotomization of continuous outcomes is especially important in medical research where cut-offs are used for diagnosis. These can be derived using different approaches: empirical approaches, where thresholds are obtained through summary statistics of measurements taken from healthy individuals; clinical approaches, which utilize a specified threshold above which symptoms and complications become more frequent; prognostic approaches, where a threshold is defined based on clinical results which may be symptom-less but carry an adverse prognosis; or, finally, operational approaches, where a threshold may be based on management and/or operational guidelines \citep{coggon2009epidemiology}. In this study, we have investigated the effects of dichotomization in the context of geostatistical modelling of disease prevalence data through a simulation study and two applications, and have proposed a likelihood-based metric to quantify the potential loss informaiton arising from this practice. All of these provided evidence that dichotomization of the data can lead to substantial loss of information for both parameter estimation and spatial prediction. We found that spatial correlation may alleviate the effects of dichotomization for parameter estimation but the increase in uncertainty and bias still remained substantially larger than those of the linear model. In particular, one of the key factors that more strongly affects the loss of accuracy and precision is the distance of the threshold from the mean of the underlying process. As such distance increases, both the bias and the MSE in the estimation of the mean component and in the spatial predictions of prevalence also increase. Another important factor is the magnitude of the noise variance $\tau^2$ relative to the signal variance $\sigma^2$. This was especially evident in the application of Section \ref{subsec:anaemia}, where a $\tau^2$ about 20 times larger than $\sigma^2$ led to the loss of the fine scale features in the spatial pattern of disease prevalence.

As shown in the application of Section \ref{subsec:anaemia}, when thresholds vary across individuals, these should be accounted for in the model for binary data, an aspect that has been ignored in previous studies of anaemia mapping. Also, this could be especially problematic if some of the covariates on which the the cut-offs are based are missing (e.g. the pregnancy status of a woman). An additional problem that arises in the context of anaemia epidemiology, is that the cut-offs described in Table \ref{tab:anaemia cutoffs} are based on guidelines from 1992 and 2001 \citep{world2011haemoglobin} which may be subject to amendment as scientific research or clinical practices evolve. 

We have only considered the case of a Gaussian distribution for the unstructured component $Z_i$. Assuming a symmetric distribution for $Z_i$ implies that, on average, misclassifications of individuals as false positives and false negatives balance out after dichotomization. However, if $Z_i$ followed a skewed distribution, this could introduce additional bias in the geostatistical model for binary data as more individuals could be misclassified as either false positives or false negatives. Hence, we expect that under these scenarios the negative effects of dichotomization on geostatistical inference would be even stronger than those shown in this study.

It is important to note that in our study we compared the performance of binary and linear geostatistical models for cut-offs that are dependent on the scale of the continuous measurement. In other cases, the $Y_i$ may follow a mixture distribution with a probability mass in zero. For example, malaria parasite density may exhibit this feature if a large proportion of the general population has not been infected and is thus clear of parasites. In this case dichotomization of the continuous outcome as $Y^*_i=1$ if $Y_i>0$ and $Y^*_i=0$ otherwise $Y_i=0$, would not lead to any loss of information.

A final remark relates to the computational burden of binary and linear geostatistical models. The likelihood function of the latter can be, most of the times, expressed in closed form, while the former requires numerical procedures based on analytical or Monte Carlo approximations of the likelihood function in order to be fitted. Hence, the increase in the computational burden is a further reason to avoid dichotomization of the data.

\section{Conclusion}
In the context of geostatistical inference, dichotomization of continuous outcomes can lead to a substantial loss of efficiency for both parameter estimation and spatial prediction. Such loss is further compounded as cut-offs used for dichotomization are further away from the mean. In addition, dichotomization can also result in the loss of fine scale features of disease prevalence, especially in the presence of a large noise to signal ratio. The findings in this study strongly support the conclusions drawn from previous studies that, whenever feasible, dichotomization should be avoided by developing models for the continuous measurements which can then be used to estimate prevalence.

\subsubsection*{Authors$\text{'}$ contributions}
IK wrote the first draft of the manuscript. 
IK and EG conducted the statistical analysis. 
IK, EG and TAA reviewed the manuscript.
TAA, MJK, and GT formed part of the Beyond Garki project team. TAA coordinated the Beyond Garki project, MJK co-designed and supervised the data collection, and GT provided technical support. 
 
\subsubsection*{Acknowledgements} 
We thank Dr. Luigi Sedda (Lancaster University) for his comments on the manuscript, and Dr Claudio Fronterr\`e (Lancaster University) for useful discussions on the computational aspects of the study.

We thank the study participants of the Beyond Garki and DHS projects, the staff of Malaria Consortium and the DHS program who were involved in the data collection, as well as the funders of the surveys presented.

IK holds a Commonwealth Scholarship Commission funded doctoral scholarship.
.
\subsubsection*{Competing interests}
The authors declare that they have no competing interests

\printbibliography
\newpage
\appendix
\setcounter{table}{0}
\renewcommand{\thetable}{A\arabic{table}}

\section{Simulation study results for sample size m=450}

\begin{table}[ht]
\centering
\caption{Bias and mean square error (in brackets) for $\tilde{\alpha}$, $\tilde{\sigma}^2$ and the estimate $\hat{\phi}$ obtained from the geostatistical models fitted the binary (B) and continuous (C) outcomes. The following are results when the number of observations, $n=450$ \label{tab:param_sim_results_n450}}
\resizebox{\columnwidth}{!}{%
\begin{tabular}{ccc|cc|cc|cc} \hline
 & & & \multicolumn{2}{|c|}{c=0} & \multicolumn{2}{|c|}{c=0.2} & \multicolumn{2}{|c}{c=0.4} \\
Parameter & $\tau^2$ & $\phi$ & B & C & B & C & B & C \\
\hline
$\tilde{\alpha}$ & 0.5 & 0.1 & -0.013 (0.090) & -0.014 (0.056) & 0.089 (0.216) & 0.017 (0.068) & 0.173 (0.180) & 0.045 (0.085)  \\
& 1 & 0.1 & -0.007 (0.06) & 0.001 (0.032) & 0.081 (0.325) & 0.019 (0.039) & 0.159 (0.054) & 0.034 (0.040)\\
& 2 & 0.1 & -0.009 (0.031) & -0.004 (0.017) & 0.073 (0.448) & 0.021 (0.019) & 0.120 (0.128) & 0.027 (0.022)\\ 
& 0.5 & 0.2 & -0.030 (0.292) & -0.029 (0.151) & 0.100 (0.382) & 0.013 (0.151) & 0.228 (0.417) & 0.023 (0.162)\\
& 1 & 0.2 & -0.014 (0.169) & -0.012 (0.077) & 0.093 (0.401) & 0.008 (0.071) & 0.175 (0.167) & 0.009 (0.079)\\
& 2 & 0.2 & -0.006 (0.093) & -0.001 (0.040) & 0.079 (0.497) & 0.011 (0.041) & 0.156 (0.169) & 0.015 (0.040) \\ \hline
$\tilde{\sigma}^2$& 0.5 & 0.1 & 0.603 (0.554) & 0.735 (4.688) & 0.624 (0.579) & 0.747 (4.983) & 0.570 (0.508) & 0.780 (4.932)\\
& 1 & 0.1 & 0.289 (0.137) & 0.543 (3.119) & 0.297 (0.136) & 0.679 (3.828) & 0.274 (0.122) & 0.508 (3.276) \\
& 2 & 0.1 &  0.125 (0.029) & 0.417 (2.781) & 0.112 (0.027) & 0.462 (2.806) & 0.107 (0.026) & 0.446 (2.966)\\ 
& 0.5 & 0.2 & 1.096 (2.059) & 0.289 (1.197) & 1.120 (2.086) & 0.387 (1.408) & 1.078 (2.015) & 0.230 (1.055) \\
& 1 & 0.2 & 0.566 (0.548) & 0.131 (0.334) & 0.565 (0.559) & 0.133 (0.267) & 0.533 (0.498) & 0.105 (0.218) \\
& 2 & 0.2 & 0.252 (0.126) & 0.100 (0.519) & 0.243 (0.119) & 0.116 (0.471) & 0.228 (0.114) & 0.075 (0.157) \\ \hline
$\phi$ & 0.5 & 0.1 & 0.031 (0.002) & 0.001 (0.001) & 0.033 (0.002) & 0.002 (0.001) & 0.031 (0.002) & 0.001 (0.001)\\
& 1 & 0.1 & 0.031 (0.002) & 0.001 (0.002) & 0.032 (0.002) & -0.002 (0.001) & 0.03 (0.002) & 0.001 (0.002)\\
& 2 & 0.1 &  0.028 (0.002) & 0.002 (0.002) & 0.026 (0.002) & 0.001 (0.002) & 0.024 (0.002) & 0.001 (0.003)\\ 
& 0.5 & 0.2 & 0.088 (0.018) & -0.017 (0.004) & 0.087 (0.017) & -0.017 (0.005) & 0.088 (0.018) & -0.011 (0.005) \\
& 1 & 0.2 & 0.095 (0.021) & -0.012 (0.007) & 0.093 (0.019) & -0.014 (0.006) & 0.090 (0.018) & -0.014 (0.006) \\
& 2 & 0.2 &  0.088 (0.020) & -0.011 (0.010) & 0.083 (0.018) & -0.015 (0.009) & 0.078 (0.017) & -0.016 (0.009) \\ 
\hline
\end{tabular}
}
\end{table}

\begin{table}[ht]
\centering
\caption{Bias and mean square error (in brackets), averaged over a 1/14 by 1/14 regular grid in $[0,2] \times [0,1]$ (hence, $m=450$), for the spatial predictions of prevalence obtained from the geostatistical models fitted to the binary (B) and continuous (C) outcomes.\label{tab:pred_sim_n450}}
\resizebox{\columnwidth}{!}{%
\begin{tabular}{cc|cc|cc|cc} \hline
 & & \multicolumn{2}{|c|}{c=0} & \multicolumn{2}{|c|}{c=0.2} & \multicolumn{2}{|c}{c=0.4} \\
 $\tau^2$ & $\phi$ & B & C & B & C & B & C \\
\hline
0.5 & 0.1 & 0.001 (0.053) & 0.001 (0.038) & 0.013 (0.053) & 0.001 (0.037) & 0.029 (0.051) & 0.001 (0.036) \\
1 & 0.1 & -0.001 (0.047) & 0.001 (0.037) & 0.019 (0.047) & -0.001 (0.037) & 0.034 (0.046) & -0.001 (0.035)\\
2 & 0.1 & -0.002 (0.037) & -0.001 (0.032) & 0.020 (0.037) & 0.001 (0.032) & 0.037 (0.036) & -0.001 (0.031)\\
0.5 & 0.2 & -0.001 (0.041) & 0.001 (0.029) & 0.012 (0.041) & 0.001 (0.029) & 0.026 (0.039) & -0.001 (0.027) \\
1 & 0.2 & 0.001 (0.036) & 0.001 (0.027) & 0.017 (0.036) & -0.001 (0.027) & 0.032 (0.035) & -0.001 (0.025) \\
2 & 0.200 & -0.001 (0.029) & 0.001 (0.022) & 0.020 (0.029) & -0.001 (0.022) & 0.037 (0.029) & -0.001 (0.021)\\ \hline
\end{tabular}
}
\end{table}
\end{document}